\documentclass[twocolumn,showpacs,preprintnumbers,amsmath,amssymb]{revtex4-1}

\usepackage{afterpage}
\usepackage[dvipdfmx]{graphicx}
\usepackage{color,bm}

\begin{document}

\title{
Iterative solutions for 
the gravitational lens equation in the strong deflection limit
} 
\author{Keita Takizawa}
\author{Hideki Asada} 
\affiliation{
Graduate School of Science and Technology, Hirosaki University,
Aomori 036-8561, Japan} 
\date{\today}

\begin{abstract} 
Two exact lens equations have been recently shown to be equivalent to each other, 
being consistent 
with the gravitational deflection angle of light 
from a source to an observer, 
both of which can be within a finite distance from a lens object 
[Phys. Rev. D 102, 064060 (2020)]. 
We examine methods for iterative solutions of 
the gravitational lens equations in the strong deflection limit. 
It has been so far unclear whether a convergent series expansion 
can be provided by 
the gravitational lens approach based on the geometrical optics 
for obtaining 
approximate solutions in the strong deflection limit 
in terms of a small offset angle. 
By using the ratio of the lens mass to the lens distance, 
we discuss a slightly different method for iterative solutions and 
behavior of the convergence. 
Finite distance effects begin at the third order in the iterative method. 
The iterative solutions in the strong deflection limit are 
estimated for  
Sgr $A^{*}$ and M87. 
These results suggest that only the linear order solution can be 
relevant with current observations, 
while the finite distance effects at the third order may be negligible 
in the Schwarzschild lens model for these astronomical objects. 
\end{abstract}


\maketitle

\section{Introduction}
Since the historical measurement by Eddington and his collaborators 
\cite{Eddington}, 
the gravitational small deflection of light has been a central subject 
in astronomy and cosmology. 
The Event Horizon Telescope (EHT) team has recently provided us 
a direct image of the immediate neighbor of the central black hole candidate 
of M87 galaxy \cite{EHT}. 
Furthermore, 
EHT has just reported measurements of linear polarizations 
around the same black hole candidate
\cite{EHT2021a} 
to infer the mass accretion rate via estimating the electron density and 
the magnetic field strength in the emitting region \cite{EHT2021b}.
Such groundbreaking events have 
increased the importance of 
the gravitational strong deflection of light 
not only in theoretical physics but also in real astronomy.

In most studies on the gravitational lens, 
a receiver and a source of light are assumed to be located 
in an asymptotic region of a spacetime. 
In this paper, the observer is referred to the receiver 
in order to avoid a confusion in notations 
between $r_0$ (the closest approach of light) 
and $r_O$ by using $r_R$. 

Based on the Gauss-Bonnet theorem \cite{GBMath} 
with using the optical metric for describing light rays, 
Ishihara et al. \cite{Ishihara2016, Ishihara2017} 
made an extension of the idea by Gibbons and Werner 
\cite{GW}, 
such that finite distance effects on the gravitational deflection of light 
for the receiver and source within a finite distance from a lens object 
can be discussed. 
Their approaches have been widely applied to a lot of spacetime models 
especially by Jusufi, Ovgun and their collaborators 
e.g. \cite{Jusufi2017a, Jusufi2017b, Jusufi2018, Ovgun2019, Ovgun2020},  
and also to the light propagation 
through a plasma medium 
by  Crisnejo and his collaborators 
\cite{Crisnejo2018, Crisnejo2019}. 
The finite-distance formulation has been extended 
to stationary and axisymmetric 
spacetimes such as Kerr solution \cite{Ono2017}, 
a rotating wormhole \cite{Ono2018} and 
a rotating global monopole with an angle deficit \cite{Ono2019}. 
See also \cite{Ono2019b} for a review on this subject. 
These works are still limited within asymptotically flat spacetimes. 

Without assuming asymptotic flatness of a spacetime,  
Takizawa et al. defined the gravitational deflection of light 
\cite{Takizawa2020a}. 
Conventional lens equations assume the infinite distance limit 
or the asymptotic flatness. 
Hence these lens equations are not compatible with 
the deflection angle of light in a finite-distance setup, 
though they can be used as an approximation. 
An exact lens equation consistent with the deflection angle of light 
for finite-distance receiver and source has been discussed by 
Bozza \cite{Bozza2008}. 
Bozza did not argue finite-distance effects, because the finite-distance formula 
for the gravitational deflection angle of light was not available at that time. 
Takizawa et al. obtained an alternative form of an exact lens equation, 
which is equivalent to the Bozza equation \cite{Bozza2008}, 
where Takizawa et al. equation becomes linear in the deflection angle. 
This lens equation for a weak deflection case is solved iteratively 
in the small angle approximations \cite{Takizawa2020b}. 

In the strong deflection limit, 
the logarithmic behavior appears in the deflection angle of light 
in the Schwarzschild spacetime 
\cite{Darwin, Bozza2002}. 
Later, Tsukamoto showed that such a logarithmic behavior is 
a rather general feature in a static and spherically symmetric spacetime 
with a photon sphere 
\cite{Tsukamoto2017, Tsukamoto2020}. 
Under certain approximations, 
conventional lens equations with the logarithmic term of 
the deflection angle of light in the strong deflection limit 
are solved by Bozza for the Schwarzschild lens
\cite{Bozza2002} 
and 
by Tsukamoto for static spherically symmetric spacetimes 
such as Ellis wormholes 
\cite{Tsukamoto2017, Tsukamoto2020}, 
where a source is assumed to be located nearly behind the lens object. 
The position angle $\theta$ of the lensed image can be split 
into $2N\pi$ and a small offset angle $\Delta\theta$, 
where $N$ is a positive integer 
(corresponding to the winding number of the light ray 
orbiting around the lens), 
and 
$\Delta\theta$ denotes a small offset angle to be determined 
by solving the lens equation. 
Indeed, the leading order solution in the strong deflection limit 
fits with the numerical solution, as shown by themselves.  

However, higher order terms in their approximation scheme have not 
been discussed in detail so far. 
It has been unclear whether a convergent series expansion 
can be provided by the 
geometrical-optics approach 
for obtaining 
approximate solutions in the strong deflection limit 
in terms of a small offset angle. 
One important reason for this is that 
the light rays in the strong deflection limit pass 
near a photon sphere, where the gradient of the effective potential 
for the photon motion diverges. 
Therefore, the Taylor series expansion in the small offset angle 
may not be well-defined. 
In particular, the 
Bozza-Tsukamoto  
method is not suitable for making a convincing argument  
on finite-distance effects.  
These 
effects are expected to appear at the next-leading order or higher orders. 
For the solar gravitational lens (which has the weak gravitational field as whole), 
Turyshev and Toth developed a theory of describing the wave-optics properties 
in the light propagation 
\cite{TT2019, TT2020, TT2021a, TT2021b}.

The main purpose of the present paper is 
to discuss, in the strong deflection limit, 
methods for iterative solutions of the lens equation 
consistent with the finite-distance configuration 
of the gravitational lens system. 
Instead of the expansion method in terms of the offset angle $\Delta\theta$, 
the present paper discusses a slightly different approach for iterative solutions 
in terms of the ratio of the lens mass to the lens distance. 
A main target of the strong deflection observation must be Sgr $A^{*}$ and M87, 
for which  
the ratio is about $10^{-11}$ and $10^{-10}$, respectively. 
Therefore, 
we examine whether the new method can lead to a well-convergent series 
expansion of the solution for such an astronomical parameter range. 
The detailed calculations and discussions are given below.

This paper is organized as follows. 
In Section II, 
the exact lens equations with finite-distance effects are briefly summarized. 
In Section III, 
we reexamine the known method for obtaining approximate solutions 
in terms of small offset angles in earlier publications 
\cite{Bozza2002, Tsukamoto2017}. 
In Section IV, 
we provide a slightly different method for iterative solutions 
with respect to the ratio of 
the lens mass to the lens distance. 
In Section V,  we 
discuss the iterative solutions with order-of-magnitude estimations 
for Sgr $A^{*}$ and M87. 
Section VI is devoted to the conclusion. 
Throughout this paper, we use the unit of $G = c = 1$.

\section{Lens equation and deflection angle of light with finite-distance effects}
\begin{figure}
\includegraphics[width=8.6cm]{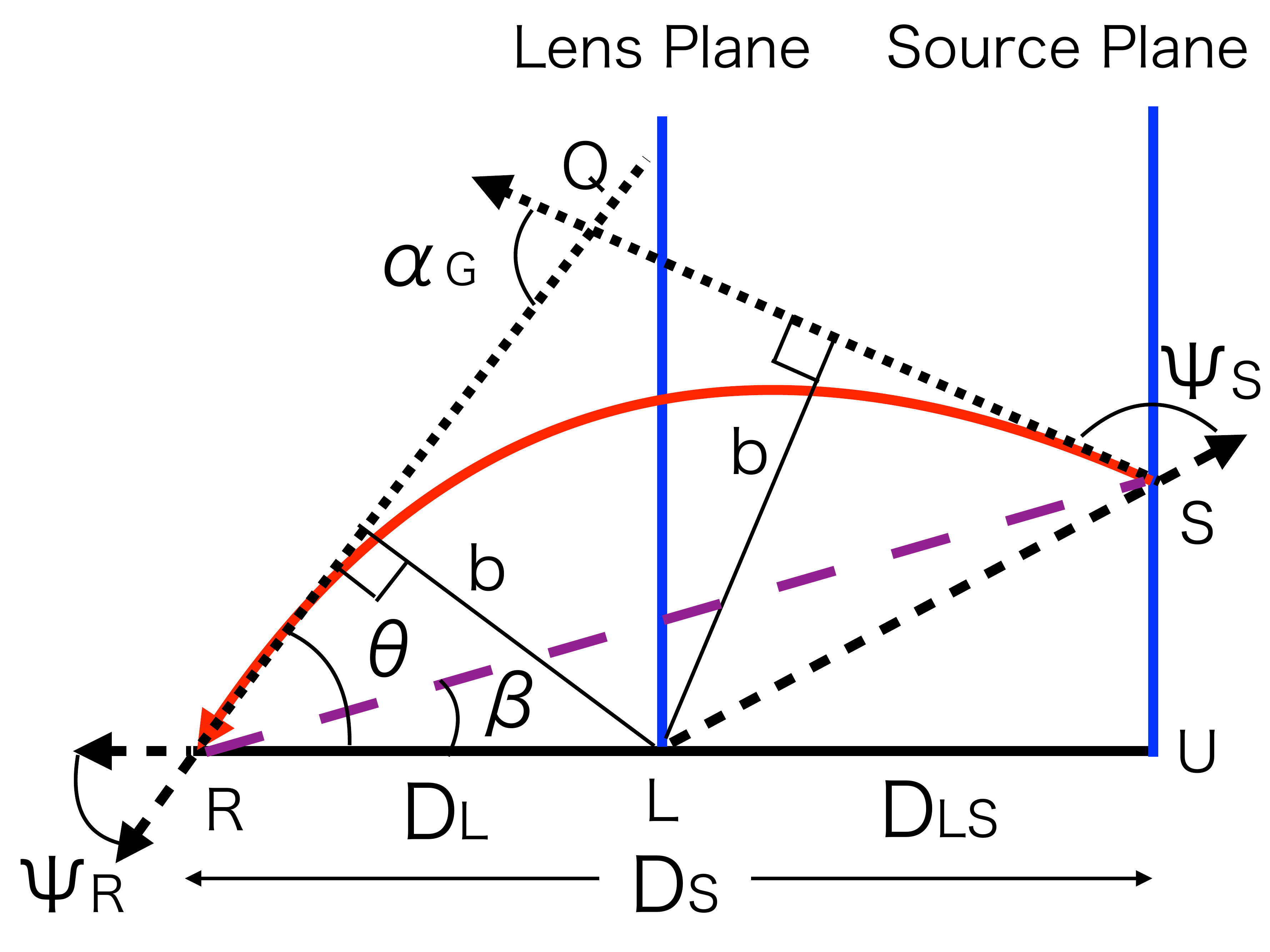}
\caption{
Geometrical configuration of a gravitational lens system. 
}
\label{fig-setup}
\end{figure}

Following Takizawa et al. \cite{Takizawa2020b}, let us summarize 
the exact lens equation that is consistent 
with the gravitational deflection angle of light 
from a source to an observer, 
both of which can be within a finite distance from a lens object. 
Lens equations in the earlier publications are based 
on additional assumptions such as a small angle approximation 
and 
a rather technical approximation that the intersection point of two tangent lines 
from the receiver and the source lies on the lens plane 
\cite{VE2000, DS}. 
Eventually, Bozza derived an improved version of the lens equation as 
\cite{Bozza2008}
\begin{align}
D_S \tan\beta 
= 
\frac{D_L \sin \theta - D_{LS} \sin(\alpha_G - \theta))}{\cos(\alpha_G - \theta)} ,
\label{lenseq-Bozza}
\end{align}
where 
$D_L$, $D_S$ and $D_{LS}$ are 
the lens distance, the source distance and the distance from the lens 
to the source, respectively, 
and 
$\beta$, $\theta$ and $\alpha_G$ denote 
the intrinsic source direction, the lensed image direction and 
the deflection angle of light, respectively. 
See Figure \ref{fig-setup} for the gravitational lens configuration.

Eq. (\ref{lenseq-Bozza}) is highly nonlinear in $\alpha_G$. 
It is thus complicated to treat this equation for an explicit form of the deflection angle 
for concrete spacetime models. 
Therefore, Takiwzawa et al. \cite{Takizawa2020b} derived 
an alternative form of the lens equation as 
\begin{align}
&\alpha_G - \theta 
- \arcsin\left(\frac{D_L}{\sqrt{(D_{LS})^2 + (D_S)^2 \tan^2 \beta}} \sin \theta \right) 
\notag\\
&+ \arctan \left( \frac{D_S}{D_{LS}} \tan\beta \right)  
\notag\\
&= 0 .  
\label{lenseq} 
\end{align}
A point is that this expression is linear in $\alpha_G$. 
Therefore, iterative calculations with Eq. (\ref{lenseq}) become much simpler than 
those using Eq. (\ref{lenseq-Bozza}). 

However, the above form does not take account of the winding number 
of the light rays. 
For the strong deflection case, the lens equation has the winding number 
in the form of 
\cite{Takizawa2020b}
\begin{align}
&\alpha_G - \theta 
- \arcsin\left(\frac{D_L}{\sqrt{(D_{LS})^2 + (D_S)^2 \tan^2 \beta}} \sin \theta \right) 
\notag\\
&+ \arctan \left( \frac{D_S}{D_{LS}} \tan\beta \right)  
\notag\\
&= 2 N \pi .  
\label{lenseq-strong} 
\end{align}

By taking account of finite-distance effects, 
Ishihara et al. obtained the deflection angle of light 
in the strong deflection limit of the Schwarzschild spacetime 
\cite{Ishihara2017}. 
\begin{align}
\alpha_I=
&\frac{2m}{b}\left[\sqrt{1-b^2u_R^2}+\sqrt{1-b^2u_S^2}-2\right]
\notag\\
&+2\log\left(\frac{12(2-\sqrt{3})r_0}{r_0-3m}\right)-\pi
\notag\\
&+O\left(\frac{m^2}{r_R{}^2},\frac{m^2}{r_S{}^2},1-\frac{3m}{r_0}\right) .
\label{alpha-strong}
\end{align}
See Appendix of Reference \cite{Ishihara2017} 
for more detail. 
In terms of the impact parameter instead of the closest approach, 
Eq. (\ref{alpha-strong}) is rearranged as 
\begin{align}
\alpha_I(\theta) =& -\frac{m}{D_L} 
\left( 1 + \frac{D_L^2}{D_{LS}^2} \right) \theta 
\notag\\
&- \ln \left( \frac{D_L \theta}{3\sqrt{3} m} - 1 \right) 
+ \ln [ 216 (7 - 4\sqrt{3}) ] - \pi . 
\label{alpha-strong-2}
\end{align}
Eq. (\ref{alpha-strong-2}) for $\alpha_I$ is substituted 
into $\alpha_G$ in Eq. (\ref{lenseq-strong}).

\section{Bozza-Tsukamoto method in the strong deflection limit}
Bozza considered the logarithmic form of the deflection angle 
in the strong deflection limit 
of Schwarzschild spacetime as 
\begin{align}
\alpha_B(\theta) =& 
- \ln \left( \frac{D_L \theta}{3\sqrt{3} m} - 1 \right) 
\notag\\
& 
+ \ln [ 216 (7 - 4\sqrt{3}) ] - \pi . 
\label{alpha-Bozza}
\end{align}
The offset in the deflection angle is defined by
\begin{align}
\Delta\alpha_N \equiv \alpha_B(\theta) - 2 N \pi . 
\label{Delta-alpha}
\end{align}

First, we define $\theta_N^0$ such that 
$\alpha_B(\theta_N^0) = 2 N \pi$. 
It is obtained as 
\begin{align}
\theta_N^0 = B (1 + e^{-C}) , 
\label{theta_N0}
\end{align}
where 
$B \equiv 3 \sqrt{3} m/D_L$ 
and 
$C \equiv (2N + 1) \pi - \ln [ 216 (7 - 4\sqrt{3}) ]$.

The offset in the image position from $\theta_N^0$ 
is defined by 
\begin{align}
\Delta\theta_N \equiv \theta - \theta_N^0 . 
\label{Delta-theta}
\end{align}

The deflection angle as Eq. (\ref{alpha-Bozza}) is  expanded 
around $\theta = \theta_N^0$, 
where $\alpha_B(\theta_N^0) = 2 N \pi$ is used. 
At the leading order, we find 
\begin{align}
\Delta\alpha_N = - \frac{e^C}{B} \Delta\theta_N .
\end{align}

The lens equation at the leading order in the strong deflection limit 
coincides with the conventional one, 
if the deflection angle $\alpha$ is replaced 
by the offset deflection $\Delta\alpha_N$  
\cite{Bozza2002}. 
\begin{align}
\beta = \theta - \frac{D_{LS}}{D_S} \Delta\alpha_N .
\end{align}
By solving this equation at the leading order in the offset angle, 
we find 
\begin{align}
\Delta\theta_N = \frac{B e^{-C} (\beta - \theta_N^0)  D_S}{D_{LS}} ,
\label{Delta-theta-Bozza}
\end{align}
which agrees with Eq. (81) of Reference \cite{Bozza2002}. 
See e.g. Section IV of Reference \cite{Bozza2002} for more detail. 

Let us discuss a convergence of the above expansion in $\Delta\theta_N$ 
at higher orders. 
We consider the Taylor expansion of $\alpha_B$ 
around $\theta = \theta_N^0$. 
\begin{align}
\alpha_B(\theta) 
= \sum_{k=1} \frac{1}{k!} \alpha_B^{(k)}(\theta_N^0) (\Delta\theta_N)^k , 
\label{alpha-Bozza-exp} 
\end{align}
where $\theta = \theta_N^0 + \Delta\theta_N$ and 
$\alpha_B(\theta_N^0) = 2 N \pi$ are used. 
By straightforward calculations, 
we can see
\begin{align}
\alpha_B^{(k)}(\theta_N^0) 
=  \frac{P}{(m/D_L)^k} , 
\label{alpha-Bozza-exp2}
\end{align}
where $P$ is a constant of $O(1)$.

Eq. (\ref{alpha-Bozza-exp2}) tells that 
the convergence radius of the series expansion by Eq. (\ref{alpha-Bozza-exp}) is 
$\sim m/D_L$. 
Eq. (\ref{alpha-Bozza-exp}) can converge if $\Delta\theta$ is 
within the convergence radius. 
Therefore, this convergence behavior may become marginal 
in the strong deflection case, because $ m/D_L \sim \theta$, 
though the leading-order solution by Bozza and Tsukamoto, 
Eq. (\ref{Delta-theta-Bozza}), is a good approximation. 
The convergence of the series might be slow, 
even if it is convergent. 
This situation can be intuitively understood because 
the photon orbit is drastically changed by even a small shift 
near the photon sphere and hence the Taylor expansion 
in $\Delta\theta_N$ 
is not convergent for such a photon orbit. 
Does there exist a convergent series expansion of 
the solutions in the strong deflection limit? 
This issue shall be discussed more quantitatively in the next section.

\section{Inverse Lens Distance Expansion}
As shown in the previous section, 
the small offset angle $\Delta\theta$ is not appropriate 
as an expansion parameter in iterative solutions. 
We should note that a small quantity 
exists in $\alpha_B$ by Eq. (\ref{alpha-Bozza}). 
It is the ratio of the lens mass to the lens distance, $m/D_L$. 
For the later convenience, we define
\begin{align}
\varepsilon \equiv \frac{m}{D_L} . 
\label{varepsilon}
\end{align}
In this section, by assuming the ratio $\varepsilon \ll 1$, 
we discuss a systematic method for iterative solutions 
for the exact lens equation in the strong deflection limit. 
Note that the deflection angle is expanded not 
in terms of the offset angle 
but in terms of $\varepsilon$. 
This expansion is expected to be convergent, because 
photon orbits depend smoothly on the lens mass $m$. 

First, following Bozza (2002), we assume that the source is located 
nearly behind the lens object. 
Then, the light ray in the strong deflection limit 
passes near the photon surface. 
Namely, $b/m = O(1)$. 
Therefore, $\beta \sim b/D_L \sim m/D_L = \varepsilon$. 
We can thus put 
$\beta = \varepsilon \beta_{(1)}$. 
The source position is given in the first place, when we solve the lens equation. 
Therefore, $\beta$ is not expanded in $\varepsilon$ 
in the iterative calculations. 
$\beta_{(1)}$ is one of input parameters that 
determine the gravitational lens configuration. 

Next, the image position is expanded as 
\begin{align}
\theta = \sum_{k=1}^{\infty} \varepsilon^k \theta_{(k)} .
\label{theta-exp}  
\end{align}

We substitute Eq. (\ref{theta-exp}) into 
$\theta$ in Eq. (\ref{lenseq-strong}) with Eq. (\ref{alpha-strong}). 
In the next section, 
we shall expand it  in terms of $\varepsilon$ 
to obtain iteratively $\theta_{(k)}$.

\section{Iterative solutions and their implications}
\subsection{Iterative solutions in terms of $\varepsilon$}
Let us begin by expansing the trigonometric functions 
in Eq. (\ref{lenseq-strong}). 
\begin{align}
& \arcsin\left(\frac{D_L}{\sqrt{(D_{LS})^2 + (D_S)^2 \tan^2 \beta}} \sin \theta \right)  
\notag\\
& = \frac{D_L}{D_{LS}} 
\left[ \varepsilon \theta_{(1)} 
+ \varepsilon^2 \theta_{(2)} 
\right.
\notag\\
&~~~~~
+ \varepsilon^3 
\left\{\theta_{(3)} 
- \frac16 \left( 1 - \frac{D_L{}^2}{D_{LS}{}^2} (\theta_{(1)})^3 \right)  
\right.
\notag\\
&~~~~~~~~~~~~~~~
\left. 
- \frac12 \frac{D_S{}^2}{D_{LS}{}^2} \theta_{(1)} (\beta_{(1)})^2 
\right\} 
\notag\\
&~~~~~
\left. 
+ O(\varepsilon^4) 
\right] , 
\label{arcsin-exp}
\end{align} 
and 
\begin{align}
& \arctan \left( \frac{D_S}{D_{LS}} \tan\beta \right)  
\notag\\
& = \frac{D_S}{D_{LS}} 
\left[ 
\varepsilon \beta_{(1)} 
+ \frac13 \varepsilon^3 
\left(
1 - \frac{D_S{}^2}{D_{LS}{}^2} 
\right) 
(\beta_{(1)})^3 
+ O(\varepsilon^5) 
\right] . 
\label{arctan-exp}
\end{align} 

The logarithmic function including $\theta$ in Eq. (\ref{alpha-strong-2}) 
is expanded as 
\begin{align}
\ln \left( \frac{D_L \theta}{3\sqrt{3} m} - 1 \right)
=& 
\ln \left( \frac{T}{3\sqrt{3}} \right) 
\notag\\
&
+ \varepsilon \frac{\theta_{(2)}}{T} 
+ \varepsilon^2 
\left\{
\frac{\theta_{(3)}}{T} 
- \frac12 \left( \frac{\theta_{(2)}}{T} \right)^2 
\right\}
\notag\\
&
+ \varepsilon^3 
\left\{
\frac{\theta_{(4)}}{T} 
- \frac{\theta_{(2)} \theta_{(3)}}{T^2} 
- \frac13 \left( \frac{\theta_{(2)}}{T} \right)^3 
\right\}
\notag\\
&
+ O(\varepsilon^4) 
\label{alpha-exp}
\end{align}
where we define 
$T \equiv \theta_{(1)} - 3\sqrt{3}$. 

By using the expansions as Eqs. (\ref{arcsin-exp}), 
(\ref{arctan-exp}) and (\ref{alpha-exp}), 
Eq. (\ref{lenseq-strong}) can be expanded in terms of $\varepsilon$. 

At the lowest order in $\varepsilon$, Eq. (\ref{lenseq-strong}) becomes 
$O(\varepsilon^0)$ as 
\begin{align}
\ln \left( \frac{T}{3\sqrt{3}} \right) = - C .
\end{align}
This equation for $\theta_{(1)}$ is solved as 
\begin{align}
\theta_{(1)} = 3\sqrt{3} (1 + e^{-C}) . 
\label{theta-1}
\end{align}

At the next order as $O(\varepsilon)$, Eq. (\ref{lenseq-strong}) becomes 
\begin{align}
\frac{\theta_{(2)}}{T} 
= \frac{D_S}{D_{LS}} (\beta_{(1)} - \theta_{(1)}) . 
\label{theta-2-pre}
\end{align}
This is solved for $\theta_{(2)}$ as 
\begin{align}
\theta_{(2)} = \frac{D_S}{D_{LS}} 
(\theta_{(1)} - 3\sqrt{3}) 
(\beta_{(1)} - \theta_{(1)}) . 
\label{theta-2} 
\end{align}

At $O(\varepsilon^2)$, Eq. (\ref{lenseq-strong}) is rearranged as 
\begin{align}
\frac{\theta_{(3)}}{T} 
= 
- \left( 1 + \frac{D_L{}^2}{D_{LS}{}^2} \right) \theta_{(1)} 
- \frac{D_S}{D_{LS}} \theta_{(2)} 
+ \frac12 \left( \frac{\theta_{(2)}}{T} \right)^2  ,
\end{align}
where $D_L + D_{LS} = D_S$ is used. 
This is solved for $\theta_{(3)}$ as 
\begin{align}
\theta_{(3)} 
=& 
-T 
\left[
\left( 1 + \frac{D_L{}^2}{D_{LS}{}^2} \right) \theta_{(1)} 
\right.
\notag\\
&~~~~~~~~
\left.
+ \frac12 \left(\frac{D_S}{D_{LS}}\right)^2 
(\beta_1 - \theta_{(1)}) 
(3 \theta_{(1)} - \beta_1 - 6\sqrt{3})
\right] . 
\label{theta-3} 
\end{align}
The first line of the right-hand side of Eq. (\ref{theta-3}) is due to 
finite-distance effects, 
while the second line appears in the strong deflection limit 
for the asymptotic receiver and source. 
In order to make a comparison, 
the third order solution in the asymptotic limit is denoted as $\theta_{(3)}^{A}$
and the finite-distance effects on the third order solution 
is denoted as $\theta_{(3)}^{F}$. 
Namely, from Eq. (\ref{theta-3}), we define 
\begin{align}
\theta_{(3)}^{A} 
&\equiv 
- \frac12 T \left(\frac{D_S}{D_{LS}}\right)^2 
(\beta_1 - \theta_{(1)}) 
(3 \theta_{(1)} - \beta_1 - 6\sqrt{3}) , 
\\
\theta_{(3)}^{F}
&\equiv
- \left( 1 + \frac{D_L{}^2}{D_{LS}{}^2} \right) T \theta_{(1)} .
\end{align}

We examine whether the series expansion in this iteration method 
is convergent. 
The above results such as Eqs. (\ref{theta-1}), (\ref{theta-2}) 
and (\ref{theta-3}) tell that 
$\theta_{(k)} \sim (D_S/D_{LS})^{k-1}$.   
The convergence radius is thus $\sim D_{LS}/D_S$. 
The series expansion by Eq. (\ref{theta-exp}) converges, 
if $\varepsilon$ is within the convergence radius, 
namely $\varepsilon < D_{LS}/D_S$. 
Substituting $D_S = D_L + D_{LS}$ into this inequality, 
we obtain 
\begin{align}
D_{LS} > \frac{m D_L}{D_L - m} ,
\label{convergence-1}
\end{align}
where we used $\varepsilon = m/D_L$. 

The present paper considers that the receiver lives far outside the horizon, 
for which $m \ll D_L$. 
In this case, the right-hand side of Eq. (\ref{convergence-1}) 
becomes $\sim m$. 
Therefore, the convergence condition Eq. (\ref{convergence-1}) is 
$D_{LS} > m$. 
Except for the particular case that the source lives 
very near the photon surface ($D_{LS} \sim m$), 
this is satisfied for $D_{LS} > m$ and Eq. (\ref{theta-exp}) converges. 
In particular, Eq. (\ref{theta-exp}) converges fast, 
because $D_{LS} \gg m$ in most of astronomical situations.

\begin{figure}
\includegraphics[width=8.6cm]{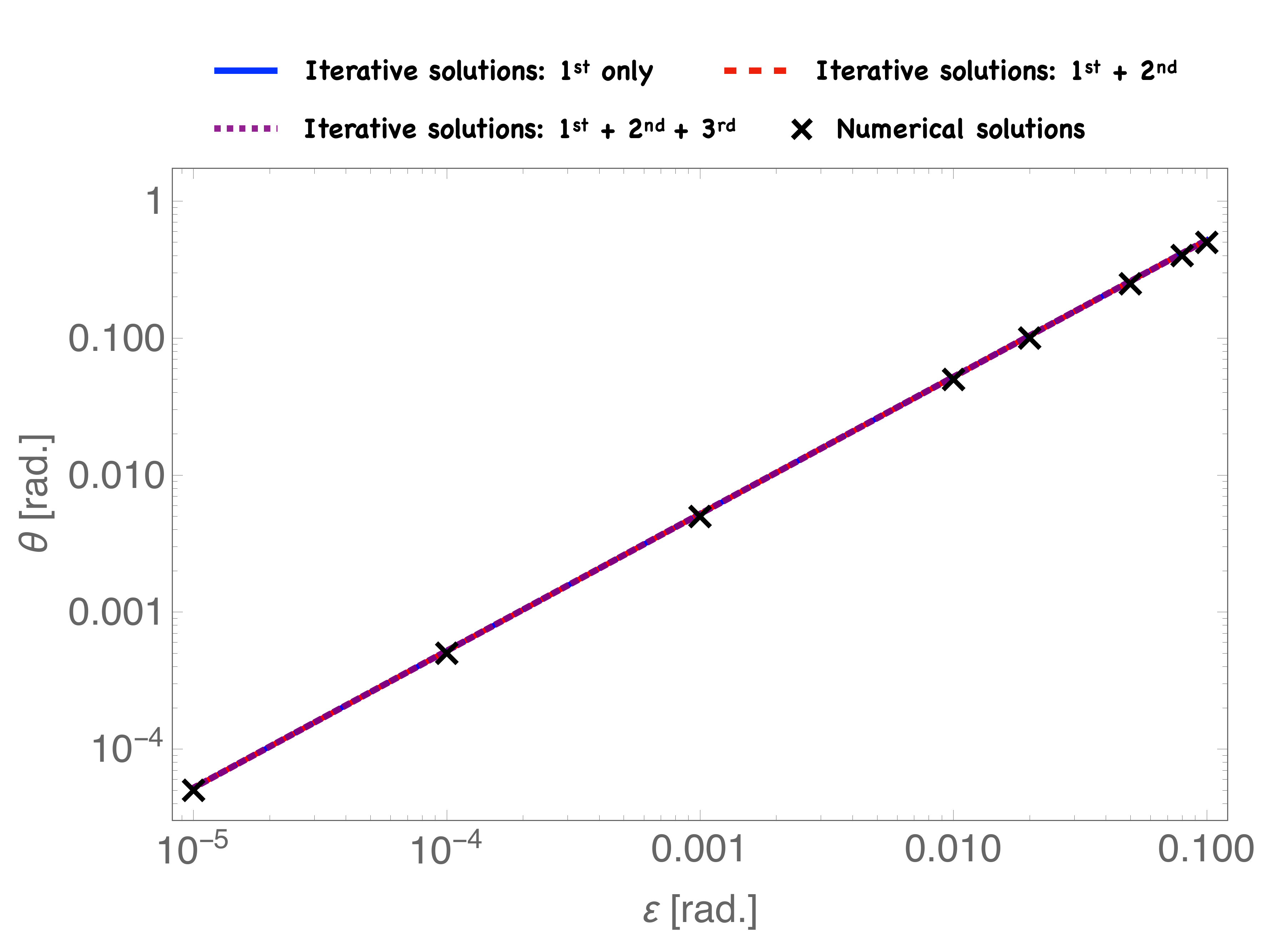}
\includegraphics[width=8.6cm]{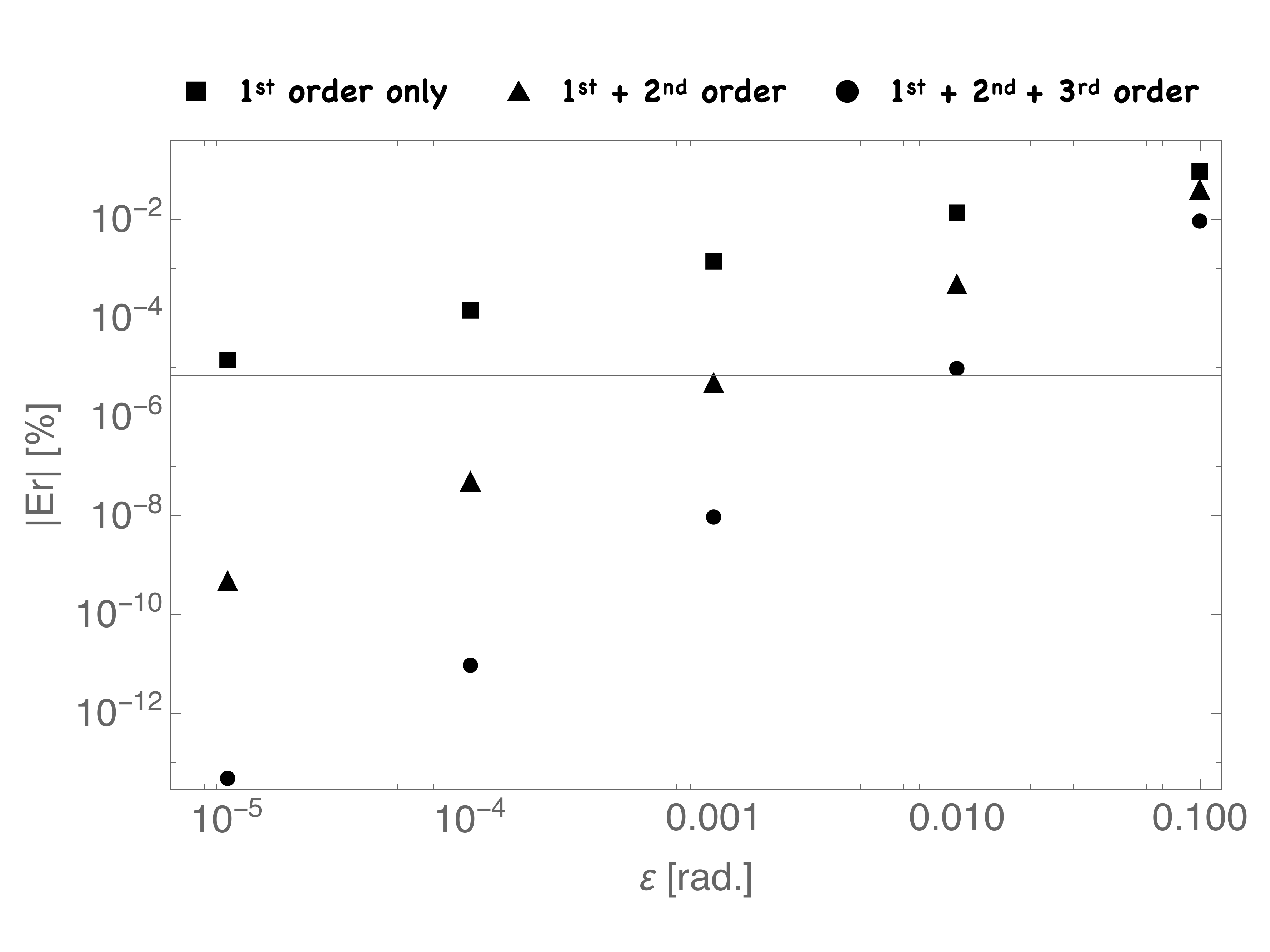}
\caption{
Convergence of iterations 
for $\beta =0$. 
Top Panel: Iterative solutions, 
where the vertical axis means the image position $\theta$. 
Bottom Panel: Relative errors, which are denoted by 
the vertical axis. 
In the two figures, the horizontal axis denotes $\varepsilon$. 
The dotted, dashed and solid curves mean 
only the first order solution, 
the solutions up to the second order 
and 
the ones up to the third order as  
$\varepsilon^3 (\theta_{(3)}^A + \theta_{(3)}^F)$ 
by using Eqs. (\ref{theta-1}), (\ref{theta-2}) and (\ref{theta-3}), respectively.  
The numerical solutions for Eq. (\ref{lenseq}) with substituting 
Eq. (\ref{alpha-strong-2}) are denoted by the cross marks. 
The three curves are overlapped 
even in the near region $\varepsilon \sim 0.1$. 
This means that the convergence is quite fast overall. }
\label{fig-convergence}
\end{figure}

Figure \ref{fig-convergence} shows a convergence of the iterative solutions. 
Overall, the iterative solutions agree well with the numerical ones. 
Importantly the convergence is very fast. 
We consider a case of $\varepsilon \sim 0.1$, 
though the present approximation should be worse in this situation. 
Even for this case, 
the linear-order solution fits very well 
with the numerical solution. 
The correction at the second order is expected to be 
$\sim \varepsilon^2 \sim 0.01$ for $\varepsilon = 0.1$, 
while the linear-order solution is apparently 
$\varepsilon \sim 0.1$. 
We may thus think that the linear-order solution with neglecting 
the second (and higher) solutions could have about ten percent errors. 
Surprisingly, however, the numerical calculations show that the error 
of the linear-order solution is only 0.085 percents even for $N=1$, 
where the error is defined as 
$|\theta - \varepsilon \theta_{(1)}|/|\theta|$ and 
$\theta$ is obtained by numerically solving Eq. (\ref{lenseq-strong}). 

Why does even the linear-order solution fit so well?  
A reason is that a factor $T$ exists in the second-order solution. 
See Eqs. (\ref{theta-2-pre}) and (\ref{theta-2}). 
$T$ can be rewritten as $T= 3 \sqrt{3} \exp(-C)$. 
We shall estimate it. 
$C$ is an increasing function of the winding number $N$. 
In the strong deflection case $N \geq1$, 
$C$ becomes the minimum ($\sim 5$) when $N=1$. 
$T$ is $\sim 5 \times 10^{-3}$ as its maximum.  
This factor $T$ is extremely small. 
This is a major reason why the second-order solution becomes extremely small. 
For its simplicity, we assume $D_L = D_{LS}$.  
The linear-order solution and the second-order oner are  
$\varepsilon \theta_{(1)} \sim 0.5$ and 
$\varepsilon^2 \theta_{(2)} \sim  -5 \times 10^{-4}$, 
where $N=1$ and $\varepsilon =0.1$.  
This implies that the relative error of the linear-order solution 
(which can be defined as 
$|\varepsilon^2 \theta_{(2)} - \varepsilon \theta_{(1)}|/|\varepsilon \theta_{(1)}|$)  
is $\sim 1 \times 10^{-3}$. 
This value is consistent with the above numerical estimation as 
$0.00085 \sim 1 \times 10^{-3}$. 

If $N=2$, $C$ is $\sim 13$, 
which leads to $T \sim 1 \times 10^{-3}$. 
This value of $T$ is smaller by two digits than that for $N=1$. 
In this way, 
through the factor $T \propto \exp (-C) \propto \exp (-2N\pi)$, 
a larger winding number $N$ significantly accelerates 
the convergence of the iterations.

\subsection{Lensing magnification and shear} 
\begin{figure}
\includegraphics[width=8.6cm]{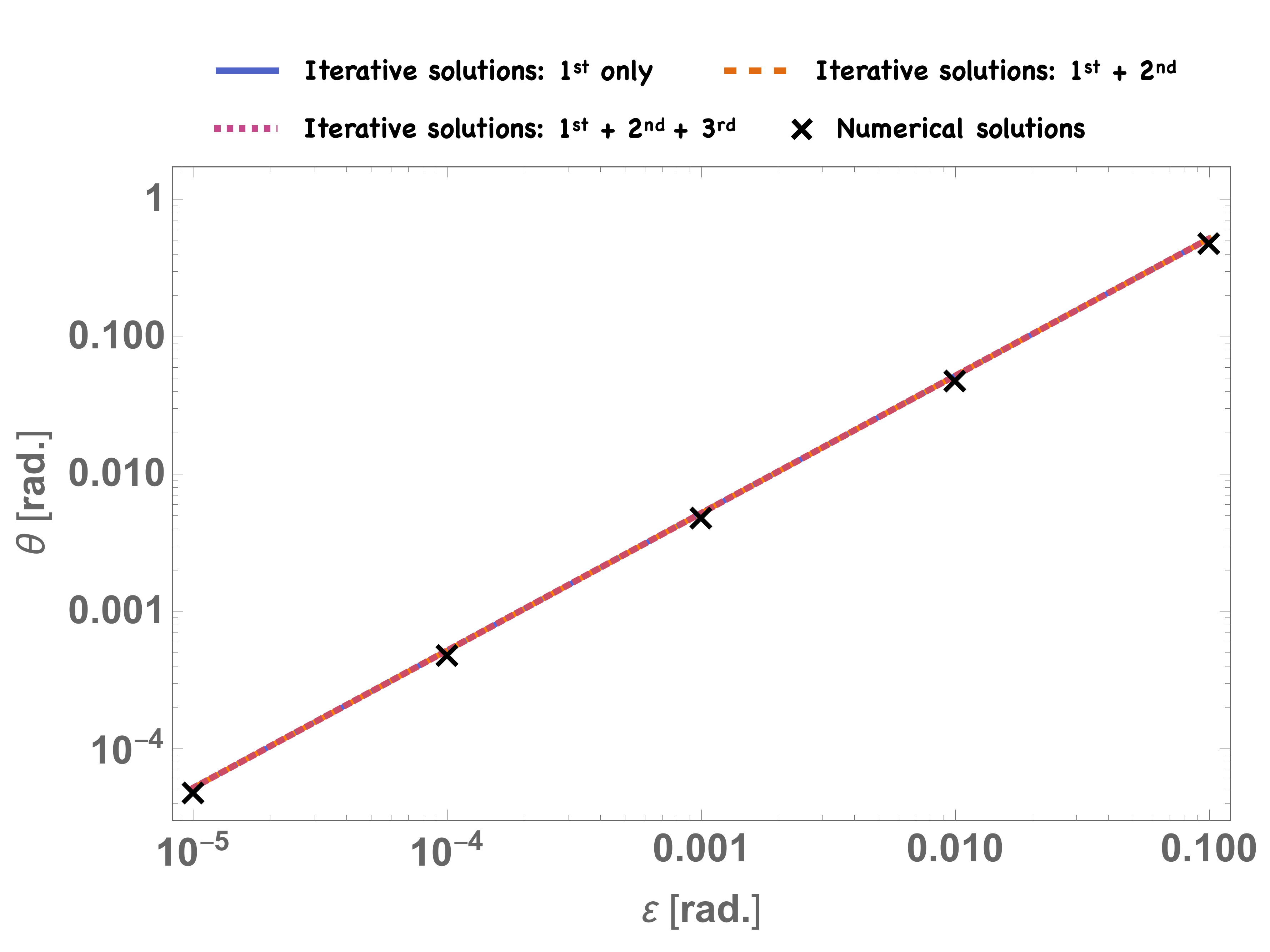}
\includegraphics[width=8.6cm]{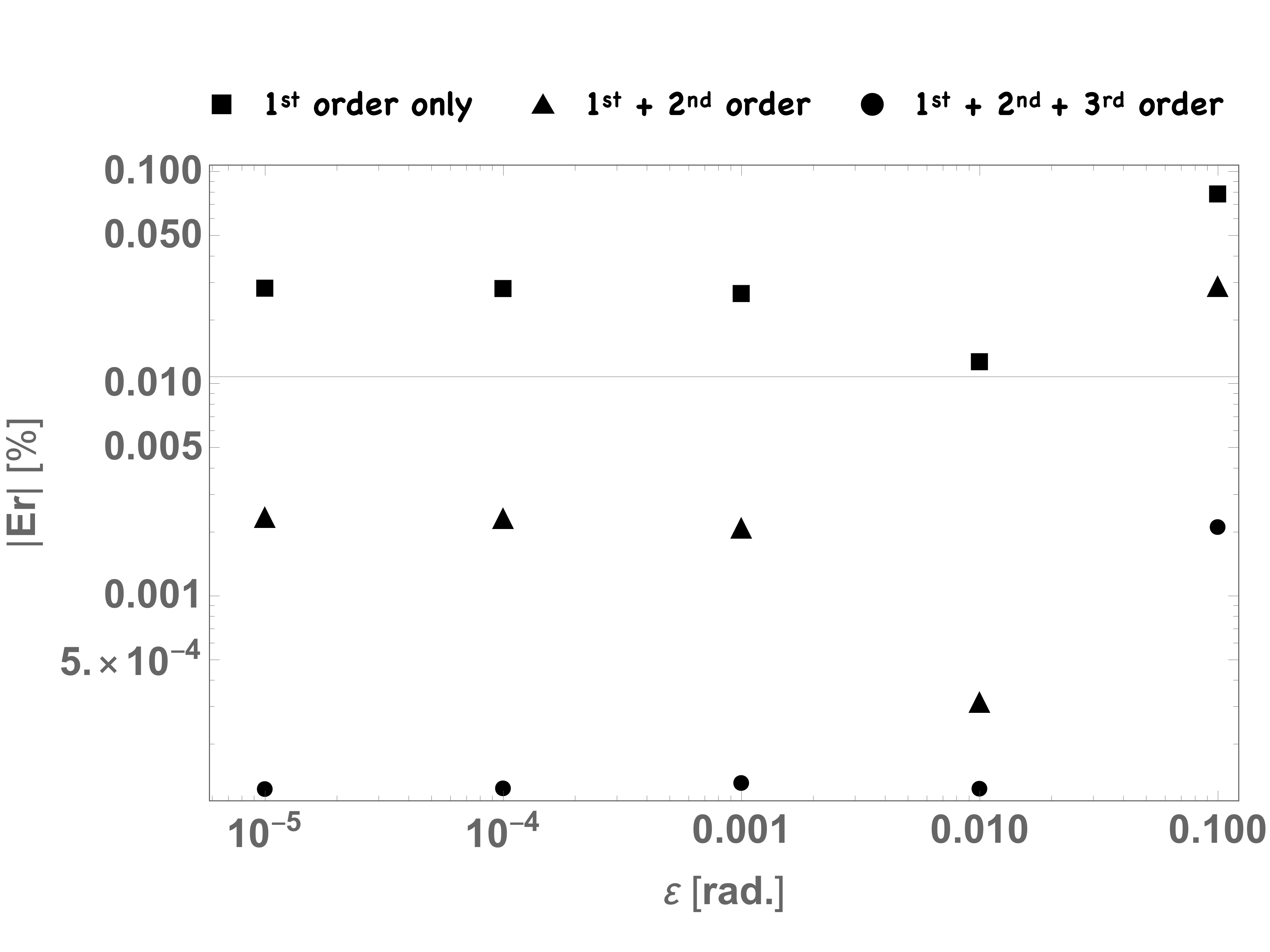}
\caption{
Iterative solutions for $\beta = 0.1$, 
which are corresponding to Figure \ref{fig-convergence} for $\beta =0$. 
Top Panel: Iterative solutions. 
Bottom Panel: Relative errors. }
\label{fig-convergence2}
\end{figure}

Higher order corrections in the image position with the exact lens equation 
may affect the brightness and the distortion of lensed images. 
In order to discuss such effects, we consider $\beta \neq 0$, 
because $\beta=0$ produces only the Einstein ring 
as a very symmetric lensed image. 
First, we make a plot for nonzero $\beta$ 
corresponding to Figure \ref{fig-convergence}. 
Figure \ref{fig-convergence2} shows the behavior of the iterative solutions 
for $\beta = 0.1$.

\begin{figure}
\includegraphics[width=8.6cm]{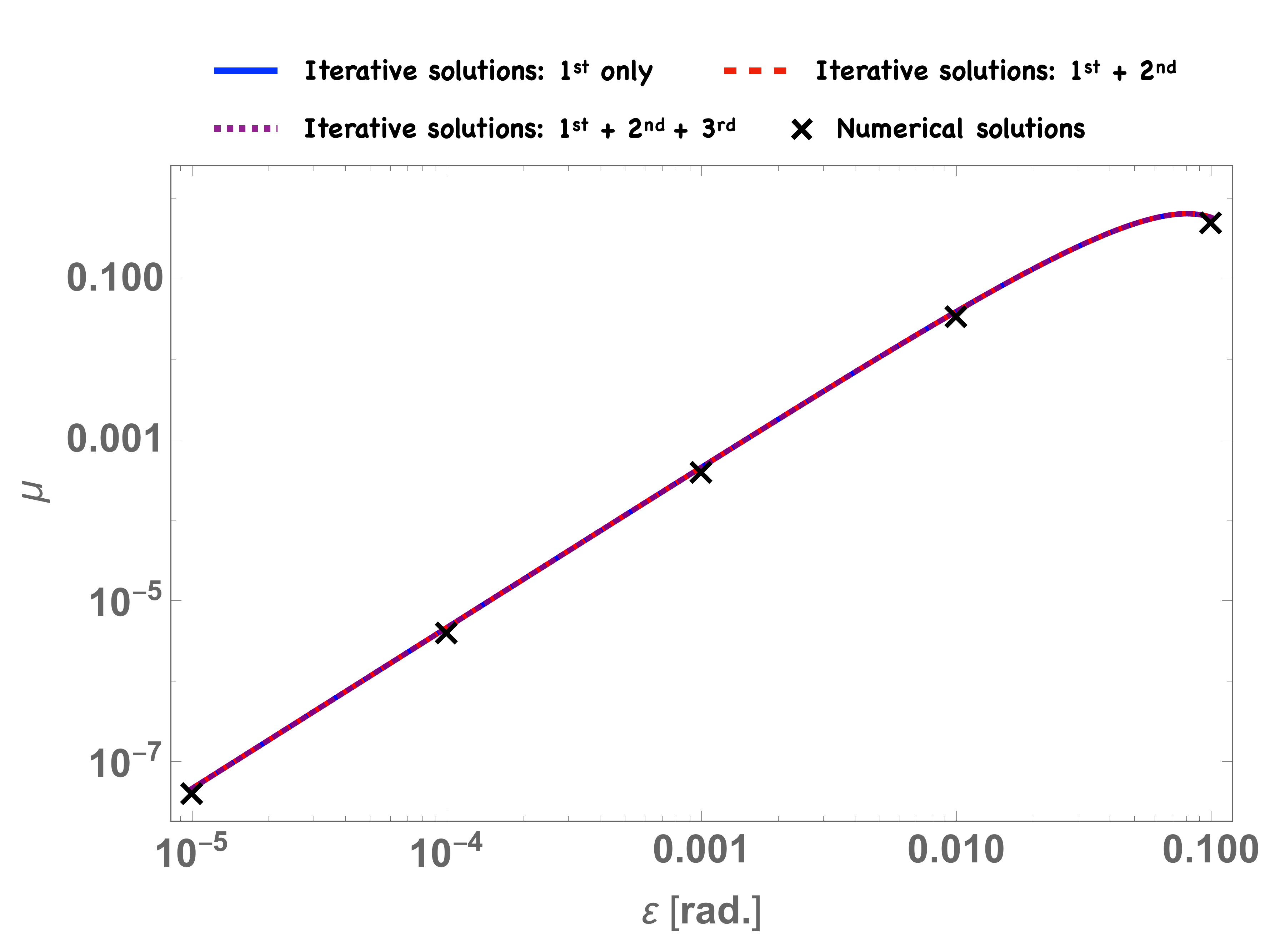}
\includegraphics[width=8.6cm]{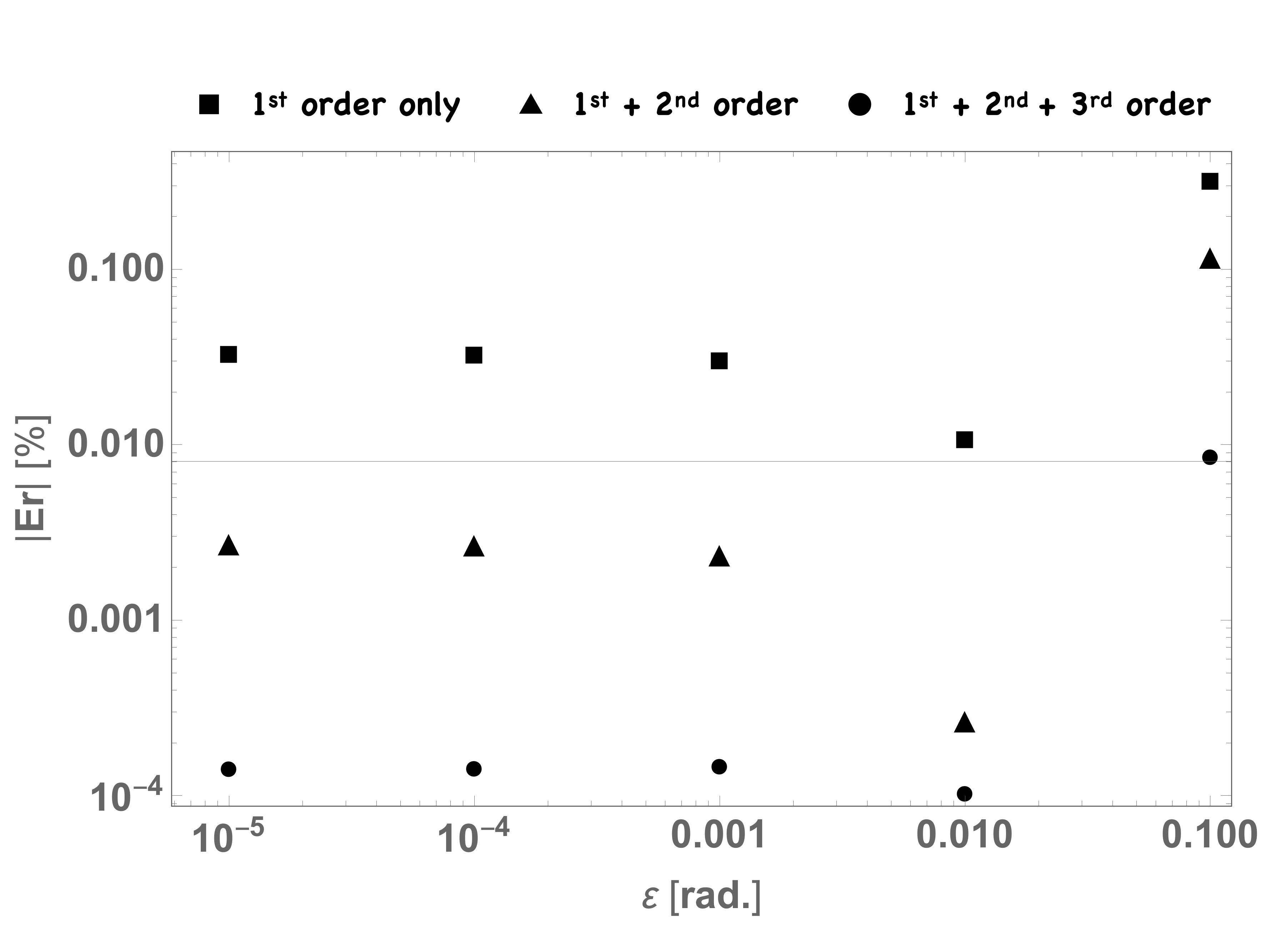}
\caption{
Lensing magnification of images for $\beta=0.1$, 
which is corresponding to Figure \ref{fig-convergence2}. 
Top Panel: Magnification factors from iterative solutions, 
where the vertical axis means the magnification factor $\mu$. 
Bottom Panel: Relative errors, which are denoted by 
the vertical axis. 
}
\label{fig-magnification}
\end{figure}

Figure \ref{fig-magnification} shows the image magnification by the iterative solutions. 
In order to calculate the magnification factor $|(\theta/\beta) (d\theta/d\beta)|$ 
for a spherical lens, we write down a useful expression of $d\theta/d\beta$ as 
\begin{align}
&\frac{d\theta}{d\beta} = 
\notag\\
& 
\frac
{D_S \cos^2 (\alpha - \theta)}
{D_L \cos^2\beta 
\left[ 
\cos\alpha + 
\dfrac{d\alpha}{d\theta} 
\sin\theta \sin(\alpha - \theta) 
- 
\dfrac{D_{LS}}{D_L}  
\left(
\dfrac{d\alpha}{d\theta} - 1 
\right)
\right]} , 
\label{dtheta} 
\end{align}
where Eq. (\ref{lenseq-strong}) (or Eq. (\ref{lenseq-Bozza}) equivalently) is used. 
The right-hand side of Eq. (\ref{dtheta}) is a function of $\theta$ 
through $\alpha = \alpha(\theta)$. 
When evaluating the magnification factor in Figure \ref{fig-magnification} 
in order to examine its convergence, 
we consider four cases; 
the first order solution, 
the solutions up to the second order, 
the ones up to the third order and the numerical ones. 
The size of the relative errors for the magnification factor 
in Figure \ref{fig-magnification} 
is consistent with the order-of-magnitude argument 
taking account of the factor $T$ 
in the similar manner to the image positions in Figure \ref{fig-convergence}.

\begin{figure}
\includegraphics[width=8.6cm]{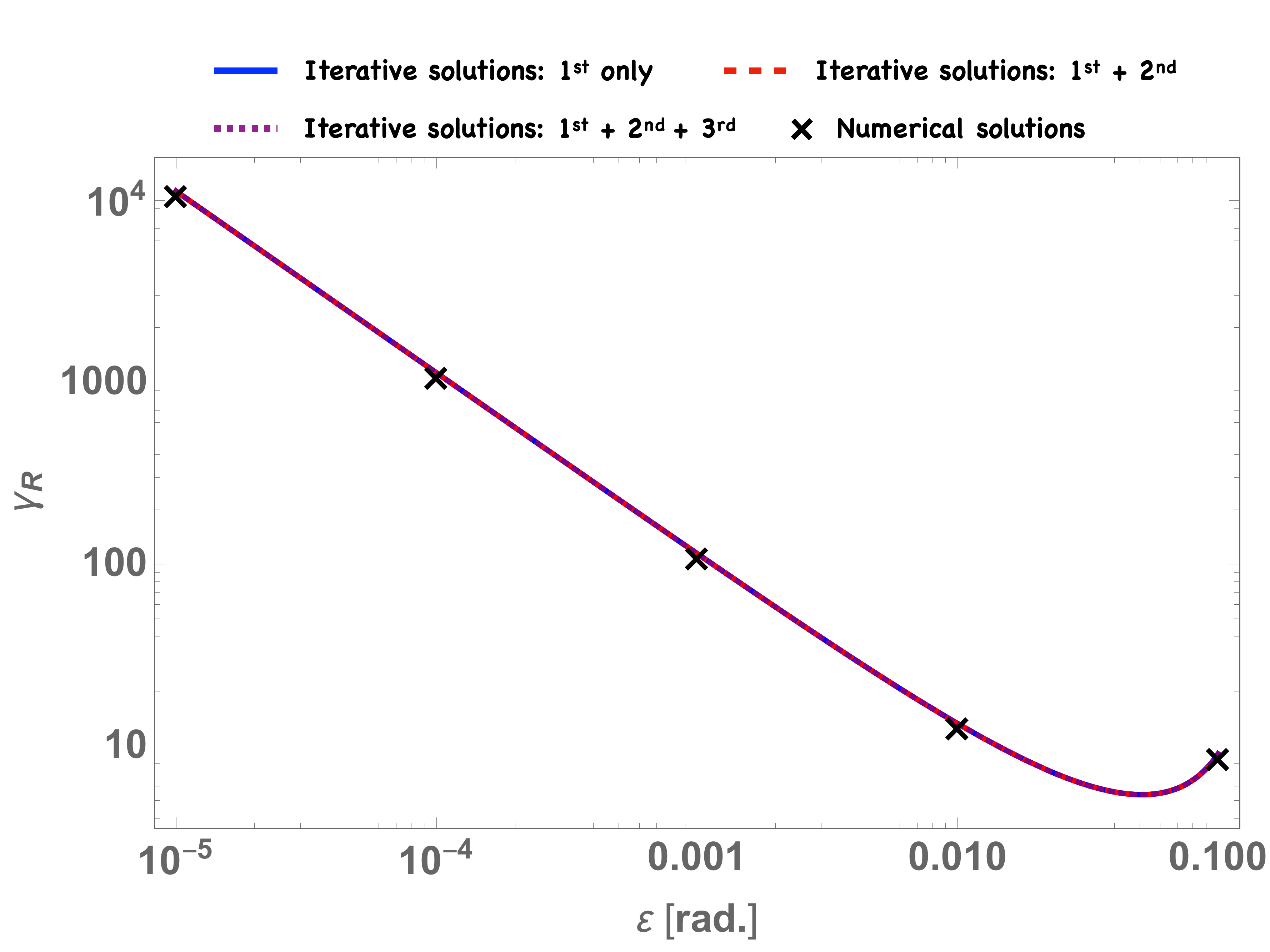}
\includegraphics[width=8.6cm]{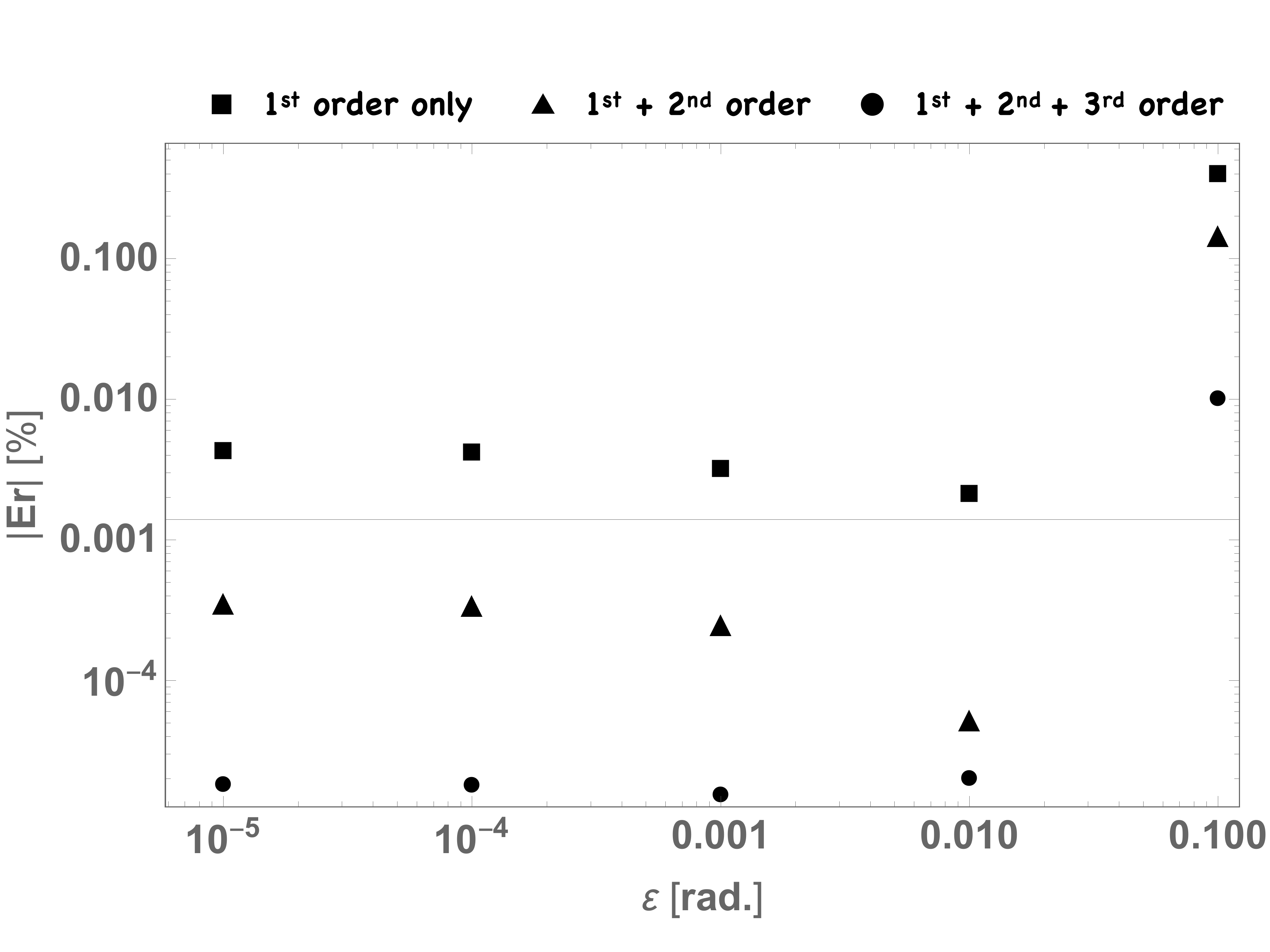}
\caption{
Radial component of the lensing shear for the image 
corresponding to Figure \ref{fig-convergence2} 
for $\beta=0.1$. 
Top Panel: Lensing shear from iterative solutions, 
where the vertical axis means the radial shear $\gamma_R$. 
Bottom Panel: Relative errors, which are denoted by 
the vertical axis. 
}
\label{fig-shearR}
\end{figure}

\begin{figure}
\includegraphics[width=8.6cm]{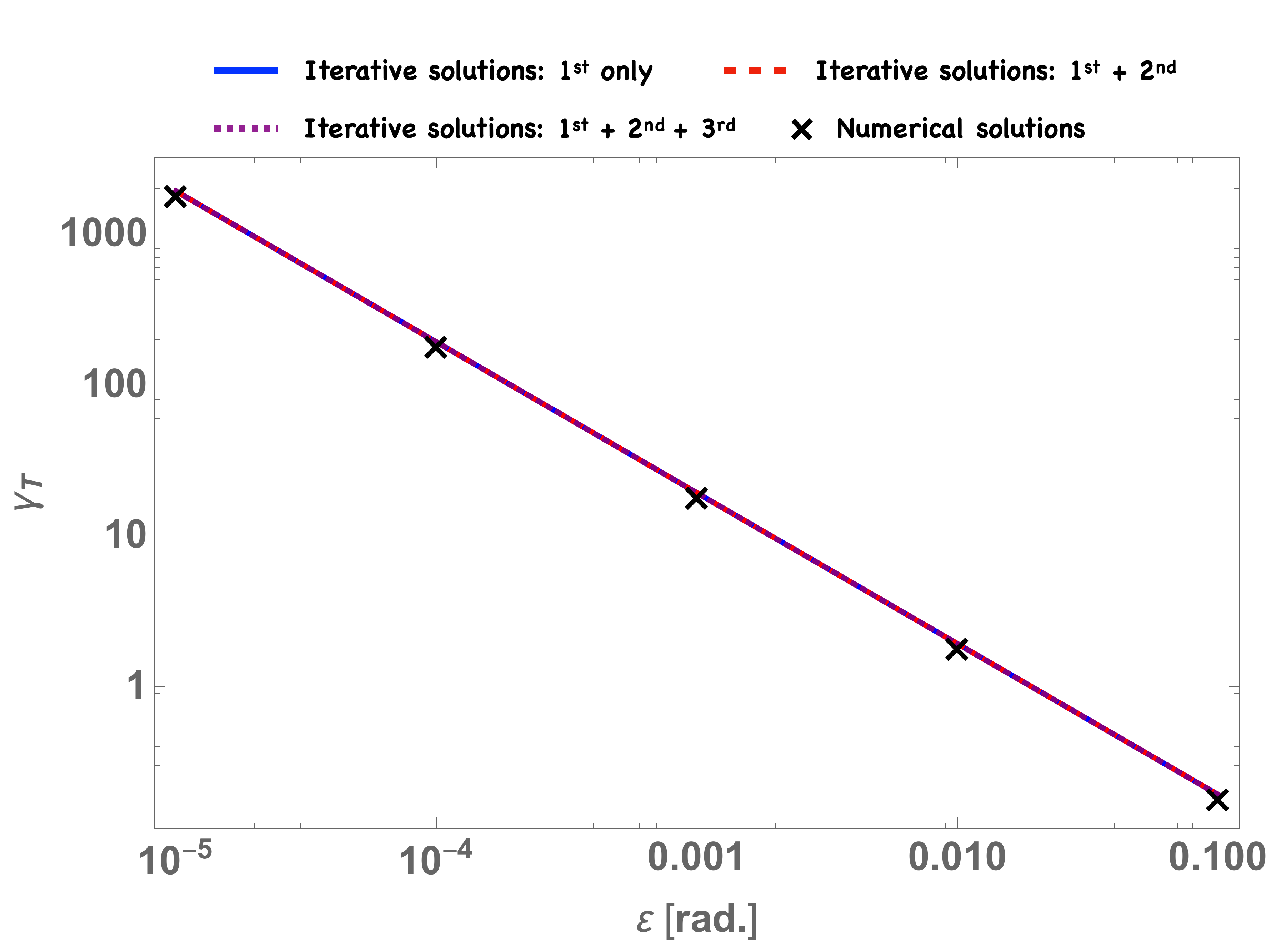}
\includegraphics[width=8.6cm]{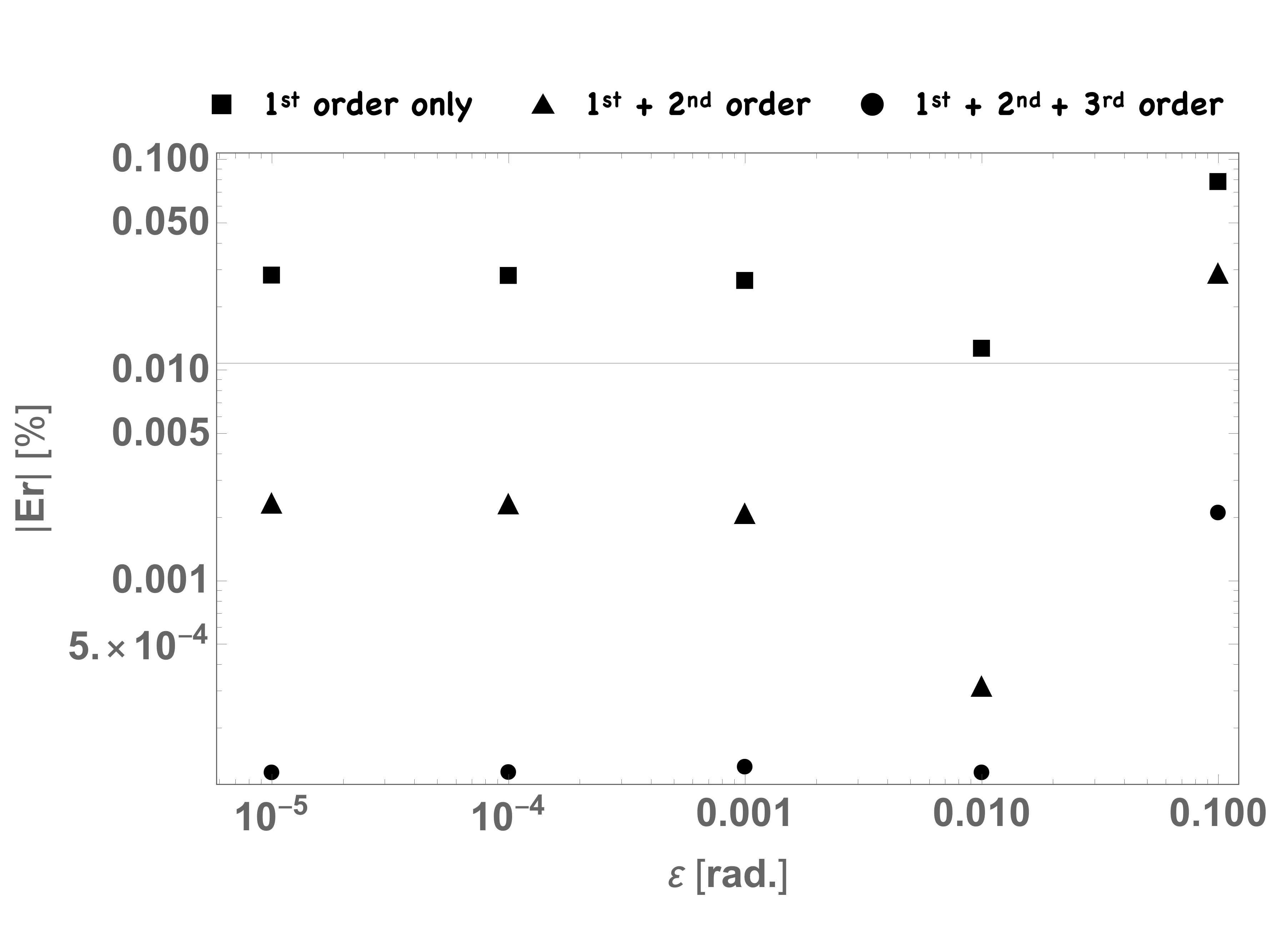}
\caption{
Tangential component of the lensing shear for the image 
corresponding to Figure \ref{fig-convergence2}  
for $\beta=0.1$. 
Top Panel: Lensing shear from iterative solutions, 
where the vertical axis means the tangential shear $\gamma_T$. 
Bottom Panel: Relative errors, which are denoted by 
the vertical axis. 
These figures correspond to Figure \ref{fig-shearR}. 
}
\label{fig-shearT}
\end{figure}

For a spherical lens, the radial shear $\lambda_R$ 
and the tangent one $\lambda_T$ become 
\cite{SEF} 
\begin{align}
\lambda_R 
&= 
\frac{d\beta}{d\theta} ,
\notag\\
\lambda_T 
&= 
\frac{\beta}{\theta} . 
\label{shear} 
\end{align}
Figures \ref{fig-shearR} and \ref{fig-shearT} show the radial shear and 
the tangential one, respectively, 
which are leading to the distortion of the lensed images. 
The size of the relative errors in these figures also 
is consistent with the order-of-magnitude discussion 
for the image positions in Figure \ref{fig-convergence}.

\subsection{
Iterative effects by some supermassive black holes}
How large are the second order solution and the third order one? 
Let us make an order-of-magnitude estimate of 
each iterative solution by assuming some supermassive black hole candidates. 
We consider two cases, 
Sgr $A^{*}$ ($m \sim 10^6 M_{\odot}$, $D_L \sim 10$ kpc) and 
M87 ($m \sim 10^{10} M_{\odot}$, $D_L \sim 10$ Mpc). 
The ratio $\varepsilon$ for Sgr $A^{*}$ and M87 is 
$10^{-11}$ and $10^{-10}$, respectively. 

See Table \ref{table-1} for Sgr $A^{*}$ and M87, 
where the image position at the leading order is $\varepsilon \theta_{(1)}$, 
the second order solution is $\varepsilon^2 \theta_{(2)}$, 
the third order solution in the asymptotic limit is $\varepsilon^3 \theta_{(3)}^{A}$
and the effects on the third order solution 
is $\varepsilon^3 \theta_{(3)}^{F}$. 
The linear order solution for Sgr $A^{*}$ and M87 
is $O(10)$ and $O(100)$ microarcseconds, respectively, 
where $D_S \sim 10^4 m$ corresponding to S2 star around Sgr $A^{*}$. 
They must be within reach of current observations. 
On the other hand, the second order and third order solutions 
for Sgr $A^{*}$ are $10^{-4}$ and $10^{-7}$ microarcseconds, respectively, 
and they for M87 are $10^{-2}$ and $10^{-6}$ microarcseconds, respectively. 
Therefore, higher order solutions are unlikely to be relevant with 
current and near-future observations.

\begin{table}
\caption{
Summary of a typical size of the iterative solutions 
for Sgr $A^{*}$ and M87, 
where $D_S \sim 10^4 m$. 
The linear order ($\varepsilon \theta_{(1)}$), 
the second order ($\varepsilon^2 \theta_{(2)}$), 
the asymptotic limit of the third order ($\varepsilon^3 \theta_{(3)}^{A}$)
and the finite-distance effects on the third order solutions 
($\varepsilon^3 \theta_{(3)}^{F}$) 
are estimated. 
The angles in this table are in radians, 
}
\begin{center}
\begin{tabular}{l|c|c}
   &  \;\;\; Sgr $A^{*}$ \;\;\; & \;\;\; M87 \;\;\; \\
\hline
$\varepsilon \theta_{(1)}$ & $10^{-10}$ & $10^{-9}$ \\
\hline
$\varepsilon^2 \theta_{(2)}$ & $10^{-15}$ & $10^{-13}$ \\
\hline
$\varepsilon^3 \theta_{(3)}^{A}$ & $10^{-18}$ & $10^{-17}$ \\
$\varepsilon^3 \theta_{(3)}^{F}$ & $10^{-18}$ & $10^{-17}$ 
\end{tabular}
\end{center}
\label{table-1}
\end{table}

\section{Conclusion}
We examined methods for iterative solutions of 
the gravitational lens equation in the strong deflection limit. 
By using the ratio of the lens mass to the lens distance, 
we proposed a method for iterative solutions and 
discussed its convergence. 
The iterative solutions in the strong deflection limit 
were 
estimated for  
Sgr $A^{*}$ and M87. 
These results suggest that only the linear order solution can be 
relevant with current observations, 
while the finite distance effects at the third order may be negligible 
in the Schwarzschild lens model for these astronomical objects. 

However, the present paper is limited within a Schwarzschild lens model. 
For instance, the asymptotic deflection by a Kerr black hole 
has been discussed iteratively in e.g. References \cite{Ibanez, AK, Edery}. 
It has been suggested that detailed measurements of black hole shadows 
will enable us to prove extra dimensions 
\cite{Banerjee}. 
In the wave optics, furthermore, 
Turyshev and Toth have recently argued that 
a small deviation of the solar gravitational lens from spherical symmetry 
makes significant effects on the caustic structure 
\cite{TT2021a, TT2021b}. 
Their method is restricted within the weak deflection case. 
Geometrical optics extensions 
to a more general class of a spherical lens or an asymmetric lens 
in the strong deflection limit  
are left as a future work.

\begin{acknowledgments}
We thank Marcus Werner and Toshiaki Ono for the useful discussions. 
We wish to thank Emanuel Gallo for the helpful comments on his recent works 
with his collaborators. 
We would like to thank Mareki Honma for the conversations 
on the EHT method and technology. 
We wish to thank Ariel Edery, Sumanta Chakraborty and Kumar Virbhadra 
for the useful comments on an earlier version of the manuscript. 
We thank Yuuiti Sendouda, Ryuichi Takahashi, Masumi Kasai, Kei Yamada, 
Hideyoshi Arakida, 
Ryunosuke Kotaki, Masashi Shinoda, and  Hideaki Suzuki 
for the useful conversations. 
This work was supported 
in part by Japan Society for the Promotion of Science (JSPS) 
Grant-in-Aid for Scientific Research, 
No. 20K03963 (H.A.),  
in part by Ministry of Education, Culture, Sports, Science, and Technology,  
No. 17H06359 (H.A.).  
\end{acknowledgments}


\begin{thebibliography}{99}
\bibitem{Eddington}
F. W. Dyson, A. S. Eddington, and C. Davidson, Phil. Trans.
R. Soc. A {\bf 220}, 291 (1920). 
\bibitem{EHT}
K. Akiyama et al. (Event Horizon Telescope Collaboration), 
Astrophys. J. {\bf 875}, L1 (2019);
Astrophys. J. {\bf 875}, L2 (2019); 
Astrophys. J. {\bf 875}, L3 (2019); 
Astrophys. J. {\bf 875}, L4 (2019); 
Astrophys. J. {\bf 875}, L5 (2019); 
Astrophys. J. {\bf 875}, L6 (2019). 
\bibitem{EHT2021a}
K. Akiyama et al. (Event Horizon Telescope Collaboration), 
Astrophys. J. {\bf 910}, L12 (2021).
\bibitem{EHT2021b}
K. Akiyama et al. (Event Horizon Telescope Collaboration), 
Astrophys. J. {\bf 910}, L13 (2021).
\bibitem{GBMath}
M. P. Do Carmo, {\it Differential Geometry of Curves and Surfaces}, 
pages 268-269, (Prentice-Hall,
New Jersey, 1976).
\bibitem{Ishihara2016}
A. Ishihara, Y. Suzuki, T. Ono, T. Kitamura and H. Asada, Phys. Rev. D {\bf 94},
 084015 (2016).
\bibitem{Ishihara2017}
A. Ishihara, Y. Suzuki, T. Ono and H. Asada, Phys. Rev. D {\bf 95},
 044017 (2017).
\bibitem{GW}
G. W. Gibbons and M. C. Werner, Class. Quantum Grav. {\bf 25}, 235009 (2008). 
\bibitem{Jusufi2017a} 
K. Jusufi, M. C. Werner, A. Banerjee, and A. Ovgun, 
Phys. Rev. D {\bf 95}, 104012 (2017). 
\bibitem{Jusufi2017b} 
K. Jusufi, A. Ovgun, and A. Banerjee, 
Phys. Rev. D {\bf 96}, 084036 (2017). 
\bibitem{Jusufi2018} 
K. Jusufi, and A. Ovgun, 
Phys. Rev. D {\bf 97}, 024042, (2018). 
\bibitem{Ovgun2019}
A. Ovgun, 
Phys. Rev. D {\bf 99}, 104075 (2019)
\bibitem{Ovgun2020}
Z. Li, and A. Ovgun, 
Phys. Rev. D {\bf 101}, 024040 (2020).
\bibitem{Crisnejo2018}
G. Crisnejo, and E. Gallo, 
Phys. Rev. D 97, 124016 (2018)
\bibitem{Crisnejo2019}
G. Crisnejo, E. Gallo, and K. Jusufi, 
Phys. Rev. D 100, 104045 (2019)
\bibitem{Ono2017}
T. Ono, A. Ishihara, and H. Asada, Phys. Rev. D {\bf 96}, 104037 (2017).
\bibitem{Ono2018}
T. Ono, A. Ishihara, and H. Asada, Phys. Rev. D {\bf 98}, 044047 (2018).
\bibitem{Ono2019}
T. Ono, A. Ishihara, and H. Asada, Phys. Rev. D {\bf 99}, 124030 (2019).
\bibitem{Ono2019b}
T. Ono, and H. Asada, Universe, 5(11), 218 (2019). 
\bibitem{Takizawa2020a}
K. Takizawa, T. Ono, and H. Asada, Phys. Rev. D {\bf 101} 104032 (2020). 
\bibitem{Bozza2008}
V. Bozza, 
Phys. Rev. D {\bf 78}, 103005 (2008). 
\bibitem{Takizawa2020b}
K. Takizawa, T. Ono, and H. Asada, 
Phys. Rev. D {\bf 102} 064060 (2020). 
\bibitem{Darwin}
C. Darwin, Proc. R. Soc. A 249, 180 (1959).
\bibitem{Bozza2002}
V. Bozza, 
Phys. Rev. D {\bf 66}, 103001 (2002).
\bibitem{Tsukamoto2017}
N. Tsukamoto, 
Phys. Rev. D {bf 95}, 064035 (2017).
\bibitem{Tsukamoto2020}
N. Tsukamoto, 
Phys. Rev. D {\bf 102}, 104029 (2020). 
\bibitem{TT2019}
S. G. Turyshev, and V. T. Toth, 
Phys. Rev. D {\bf 100}, 084018 (2019). 
\bibitem{TT2020}
S. G. Turyshev, and V. T. Toth, 
Phys. Rev. D {\bf 101}, 044025 (2020). 
\bibitem{TT2021a}
S. G. Turyshev, and V. T. Toth, 
Phys. Rev. D {\bf 103}, 064076 (2021).
\bibitem{TT2021b}
S. G. Turyshev, and V. T. Toth, 
arXiv:2103.06955.
\bibitem{VE2000}
K. S. Virbhadra, and G. F. R. Ellis,
Phys. Rev. D {\bf 62}, 084003 (2000).
\bibitem{DS}
M. P.  Dabrowski, and F. E. Schunck, 
Astrophys. J. {\bf 535}, 316 (2000). 
\bibitem{SEF}
P. Schneider, J. Ehlers, and E. E. Falco, 
{\it Gravitational Lenses} 
(Springer, NY, 1992). 
\bibitem{Ibanez}
J. Ibanez, 
Astron. Astrophys. {\bf 124}, 175 (1983). 
\bibitem{AK}
H. Asada, and M. Kasai, 
Prog. Theor. Phys. {\bf 104}, 95 (2000). 
\bibitem{Edery}
A. Edery, and J. Godin, 
Gen. Rel. Grav. {\bf 38}, 1715 (2006).  
\bibitem{Banerjee}
I. Banerjee, S. Chakraborty, and S. SenGupta, 
Phys. Rev. D {\bf 101}, 041301 (2020). 
\end{thebibliography}
\end{document}